\documentclass[preprint,aps,pra,showpacs,floatfix]{revtex4}
\usepackage{graphicx}
\usepackage{times}
\usepackage{nicefrac}
\usepackage{amsmath}
\usepackage{amsfonts}
\usepackage{amssymb}
\usepackage{amsthm}
\usepackage{epsf}
\usepackage{bm}
\usepackage{bbm}
\usepackage{times}

\usepackage{dcolumn}

\newcommand{\balpha}{{\mbox{\boldmath$\alpha$}}}

\newcommand{\be}{\begin{eqnarray}}
\newcommand{\ee}{\end{eqnarray}}
\newcommand{\la}{\langle}
\newcommand{\ra}{\rangle}

\newcommand{\bfx}{{\bf x}}
\newcommand{\bfy}{{\bf y}}
\newcommand{\bfk}{{\bf k}}
\newcommand{\bfr}{{\bf r}}
\newcommand{\br}{{\bf r}}
\newcommand{\bd}{{\bf d}}

\newcommand{\veps}{\varepsilon}

\newcommand{\beps}{{\mbox{\boldmath$\epsilon$}}}

\newcommand{\bmu}{{\mbox{\boldmath$\mu$}}}

\begin{document}

\title{Parity nonconservation effect
with laser-induced $2^3S_1$- $2^1S_0$ transition in heavy heliumlike ions}

\author{V. M. Shabaev,$^{1,2}$ A. V. Volotka,$^{1,3}$
C. Kozhuharov,$^{2}$ G. Plunien,$^{3}$ and Th. St\"ohlker$^{2,4,5}$}

\affiliation {$^{1}$Department of Physics, St.Petersburg State University,
Ulianovskaya 1, Petrodvorets, St.Petersburg 198504, Russia\\
$^{2}$Gesellschaft f\"{u}r Schwerionenforschung,
Planckstrasse 1, D-64291 Darmstadt, Germany\\
$^3$Institut f\"ur Theoretische Physik, Technische Universit\"at Dresden,
Mommsenstrasse 13, D-01062 Dresden, Germany\\
$^4$Physikalisches Institut, Universit\"at Heidelberg,
Philosophenweg 12, D-69120 Heidelberg, Germany\\
$^5$Helmholtz-Institut Jena, D-07743 Jena, Germany}

\begin{abstract}
The parity nonconservation (PNC) effect on the laser-induced
$2^3S_1$- $2^1S_0$ transition in heavy heliumlike
ions is considered. A simple analytical formula for the
PNC correction to the cross section is derived for the case,
when the opposite-parity $2^1S_0$ and  $2^3P_0$ states are
almost degenerate and, therefore, the PNC effect is strongly
enhanced. Numerical results are presented for heliumlike
gadolinium and thorium, which seem most promising
candidates for such kind of experiments. In both Gd and Th cases
the photon energy required  will be anticipated with
a high-energy laser built at GSI. Alternatively, it can be gained
with ultraviolet lasers utilizing relativistic Doppler tuning at FAIR
facilities in Darmstadt.
\end{abstract}
\maketitle

\section{Introduction}

Measurements of parity nonconservation (PNC) effects with
heavy few-electron ions can provide new opportunities
for tests of the Standard Model at low-energy regime.
This is mainly due to the fact that, in contrast to
neutral atoms (see Refs.~\cite{dzu89,blu00,dzu95,saf00,der01,koz01,dzu02,gin04,sha05,por09}),
in highly charged ions the electron-correlation effects,
being suppressed  by a factor $1/Z$ ($Z$ is the nuclear charge number),
can be accounted for by perturbation theory to a very high
accuracy. The simple atomic structure of such ions
allows also one to calculate the QED contributions
to the required accuracy.

PNC experiments with highly charged ions were first discussed in Ref.~\cite{gor74}.
There it was proposed to use close
opposite-parity levels  $2^1S_0$ and  $2^3P_1$ in He-like ions for
$Z\approx 6$ and  $Z\approx 29$, where the PNC effect is strongly enhanced.
Later, various scenarios for PNC experiments with heavy H- and He-like ions
were considered in a number of papers
\cite{sch89,opp91,kar92,dun96,zol97,lab01,pin93,gri05,lab07,mai09}.
In particular, in Ref.~\cite{opp91} it was proposed to study the induced
$2^3S_1$ - $2^1S_0$ transition in He-like ions with $Z\approx 6$
in the presence of electric
and magnetic fields. Possibilities to investigate PNC effects in H-like ions
at high-energy ion storage rings utilizing relativistic Doppler tuning and laser cooling
were considered in Ref.  \cite{zol97}.
Most of the works \cite{sch89,kar92,dun96,lab01,mai09} exploited, however,
the near-degeneracy of the $2^1S_0$ and  $2^3P_0$ states in He-like ions at
$Z\approx 64$ and  $Z\approx 90$.  For overviews of the
schemes suggested we refer to Refs.~\cite{lab07,mai09}.

In the present paper, we evaluate the PNC effect on
the laser-induced $2^3S_1$- $2^1S_0$ transition in heavy heliumlike ions
nearby $Z=64$ (transition energy of about 114 eV)
and $Z=90$ (transition energy of about 240 eV),
where the PNC effect is strongly enhanced.
Such experiments seem to be feasible in near future
in view of recent developments in high-energy lasers for heavy ion
experiments (PHELIX project) \cite{zim08,Bag09}.
As an alternative, one may consider employment of relativistic
Doppler tuning  at FAIR facilities in Darmstadt \cite{Bac00,Sto06}.
With ion energies up to 10.7 GeV/u, as anticipated at the FAIR facilities,
the Doppler effect can be utilized for tuning ultraviolet laser light
with photon energies in the range from 4 to 10 eV in resonance with
the transition energies under consideration.

The paper is organized as follows. In section \ref{sec2}, the basic formulas
for the $2^3S_1$- $2^1S_0$ transition amplitude are presented.
The admixture of the opposite-parity states $2^1S_0$ and $2^3P_0$
is taken into account and, as a result, the PNC correction to the cross
section is derived. It is shown, that accounting for the first-order 
interelectronic-interaction and
QED corrections in the velocity gauge can be easily done 
within the zeroth-order approximation in the length gauge.
In section \ref{sec3}, numerical results for the PNC correction
in heliumlike gadolinium and thorium are presented
and possible scenarios for experiments are discussed.

Relativistic units ($\hbar=c=1$) and the Heaviside charge
unit [$\alpha = e^2/(4\pi)$, $e<0$] are used throughout
the paper.

\section{Basic formulas}
\label{sec2}

We consider the absorption of a photon with energy
$\omega \approx E_{2^1S_0}-E_{2^3S_1}$ and circular polarization
$\lambda =\pm 1$ by a heavy heliumlike ion being initially prepared in
the $2^3S_1$ state. If the weak electron-nucleus interaction is ignored,
the absorption cross section is completely
determined by the magnetic-dipole transition amplitude.
For such a transition  the interelectronic-interaction effects
are suppressed by a factor $1/Z$ and, to zeroth order, we assume that the
electrons interact only with the Coulomb field of the nucleus.
Then,
the wave functions
of the initial ($2^3S_1$)  and the final ($2^1S_0$) state are given by
    \begin{eqnarray}
u_{JM}(\bfx_1,\bfx_2) = \frac{1}{\sqrt{2}}\sum_{m_1,m_2} C^{JM}_{j_1\,m_1,j_2\,m_2}
[\psi_{j_1\,m_1}(\bfx_1)\psi_{j_2\,m_2}(\bfx_2)
-\psi_{j_1\,m_1}(\bfx_2)\psi_{j_2\,m_2}(\bfx_1)]
\,,
 \label{wf_f}
    \end{eqnarray}
where $\psi_{j_1\,m_1}(\bfx)$ is the one-electron $1s$ wave function,
$\psi_{j_2\,m_2}(\bfx)$ is the one-electron $2s$
wave function, and  $C^{JM}_{j_1\,m_1,j_2\,m_2}$ is the Clebsch-Gordan
coefficient.
In what follows, we assume that the laser spectral width and the
width due to a finite ion-laser interaction time can be neglected.
If, for a moment, we further neglect the width of the initial
state, the
cross section in the resonant approximation is given by
(see, e.g., Refs.~\cite{ber82,sha02,bud04})
    \begin{eqnarray}
{\sigma} ={(2\pi)^3}\frac{\Gamma_b |\la b|[R(1)+R(2)]|a\ra|^2}{(E_a+\omega-E_b)^2
+\Gamma_b^2/4}
\,.
 \label{cross}
    \end{eqnarray}
Here $|a\rangle \equiv |2^3S_1\rangle$ and
$|b\rangle \equiv |2^1S_0\rangle$ are the initial and final states, respectively,
$\Gamma_b$ is the width of the final state, and $R(i)$ is the transition operator
acting on variables of the $i$th electron.
In the transverse gauge, $R =-e\balpha\cdot {\bf A}$, where
    \begin{eqnarray}
{\bf A}(\bfx) = \frac{\beps \exp{(i\bfk\cdot \bfx)}}{\sqrt{2\omega (2\pi)^3}}
\,
 \label{photon}
    \end{eqnarray}
is the wave function of the absorbed photon and $\balpha $ is the vector incorporating
the Dirac matrices.
In order to account for the width of the initial state $a$ in Eq. (\ref{cross}),
we simply replace  $E_a\rightarrow E$ and
    \begin{eqnarray}
|a\ra \la a| \rightarrow \int dE\frac{\Gamma_a/(2\pi)}{(E-E_a)^2
+\Gamma_a^2/4}|a\ra \la a|
\,.
 \label{width_a}
    \end{eqnarray}
In a more rigorous approach, one should consider the preparation of the state  $a$
as a part of the whole process \cite{sha02}.
With the substitution (\ref{width_a}), we get
    \begin{eqnarray}
{\sigma} ={(2\pi)^3} \int dE \frac{\Gamma_b \Gamma_a |\la b|[R(1)+R(2)]|a\ra|^2}
{2\pi [(E+\omega-E_b)^2
+\Gamma_b^2/4] [(E-E_a)^2
+\Gamma_a^2/4]}
\,.
 \label{cross1}
    \end{eqnarray}
Integrating over $E$, we obtain
    \begin{eqnarray}
{\sigma} ={(2\pi)^3} \frac{\Gamma_a + \Gamma_b}{[\omega-(E_b-E_a)]^2
+(\Gamma_a+\Gamma_b)^2/4}
 |\la b|[R(1)+R(2)]|a\ra|^2
\,.
 \label{cross2}
    \end{eqnarray}
In the resonance case, $\omega = E_b-E_a$, we have
    \begin{eqnarray}
{\sigma} =4{(2\pi)^3} \frac{|\la b|[R(1)+R(2)]|a\ra|^2}{\Gamma_a + \Gamma_b}
\,.
 \label{cross3}
    \end{eqnarray}
Finally, averaging over the angular momentum projection of the initial state, we
obtain
    \begin{eqnarray}
{\sigma} =4{(2\pi)^3} \frac{1}{2J_a +1}\sum_{M_a}
\frac{|\la b|[R(1)+R(2)]|a\ra|^2}{\Gamma_a + \Gamma_b}
\,.
 \label{cross4}
    \end{eqnarray}
In what follows, due to smallness of the transition energy, we can write
\be
 R =-e\balpha\cdot {\bf A}=
-e(\balpha\cdot\beps)
 \frac{ \exp{(i\bfk\cdot \bfx)}}{\sqrt{2\omega (2\pi)^3}}
\approx
-e \frac{(\balpha\cdot\beps) }{\sqrt{2\omega (2\pi)^3}}
(1+i\bfk\cdot\bfx)
\,.
\ee
For the transition $J_a=1 \rightarrow J_b=0$
we can restrict to the dipole approximation and, therefore,
represent the transition operator as the sum
\be
R=R_{\rm e}+R_{\rm m}\,,
\ee
where
\be
R_{\rm e} = -e \frac{(\beps\cdot\balpha) }{\sqrt{2\omega (2\pi)^3}}\,
\ee
is the electric-dipole transition operator in the velocity gauge,
\be
R_{\rm m} = i \frac{([\beps\times \bfk]\cdot\bmu)}{\sqrt{2\omega (2\pi)^3}}\,
\ee
is the magnetic-dipole transition operator, and $\bmu =(e/2)[\bfx\times\balpha]$
is the operator of the magnetic moment of electron. If we neglect the weak interaction,
the $2^3S_1$- $2^1S_0$ transition amplitude is
the pure magnetic-dipole one. Then, evaluating the matrix elements in
Eq. (\ref{cross4}), we obtain
    \begin{eqnarray}
{\sigma_0^{(2^3S_1\rightarrow 2^1S_0)}} =\frac{1}{9}\frac{\omega}{\Gamma_{2^3S_1} + \Gamma_{2^1S_0}}
|\la 2s||\bmu||2s\ra - \la 1s||\bmu||1s\ra|^2
\,,
 \label{cross40}
    \end{eqnarray}
where $ \la ns||\bmu||ns\ra $ is the reduced matrix element of the magnetic-dipole-moment
operator and the subscript ``0'' stays for the zeroth-order approximation.

To account for the weak interaction we have to first modify the wave function
of the $2^1S_0$ state due to the admixture of the $2^3P_0$ state:
    \begin{eqnarray}
 \label{wf_pnc1}
|2^1S_0\ra \rightarrow |2^1S_0\ra +\frac{\la 2^3P_0|[H_W(1)+H_W(2)]|2^1S_0\ra}{E_{2^1S_0}-
E_{2^3P_0}}|2^3P_0\ra\,.
    \end{eqnarray}
Here
\be \label{h_w}
H_W=-(G_F/\sqrt{8})Q_W \rho_{N}(r)\gamma_5
\ee
is spin-independent part of the effective nuclear weak-interaction Hamiltonian \cite{khr91}.
$G_F$ denotes the Fermi constant,  $Q_W \approx -N + Z(1-4{\rm sin}^2\theta_W)$
is the weak charge of the nucleus (which is related to the Weinberg angle $\theta_W$),
$\gamma_5$ is the Dirac matrix, and $\rho_{N}$  is the effective nuclear weak-charge
density normalized to unity. A simple evaluation of the weak-interaction matrix
element yields
\be \label{h_w_m}
\la 2^3P_0|[H_W(1)+H_W(2)]|2^1S_0\ra&=& \la 2p_{1/2}|H_W|2s\ra
\nonumber\\
&=&
i\frac{G_F}{2\sqrt{2}}Q_W\int_{0}^{\infty}dr\,r^2
\rho_N(r)[g_{2p_{1/2}}f_{2s}-f_{2p_{1/2}}g_{2s}]\,,
\ee
where the large and small radial components
of the Dirac wave function
are defined by
\be
\psi_{n\kappa m}({\bf r})=
\left(\begin{array}{c}
g_{n\kappa}(r)\Omega_{\kappa m}({\bf n})\\
if_{n\kappa}(r)\Omega_{-\kappa m}({\bf n})
\end{array}\right)\;
\ee
and $\kappa=(-1)^{j+l+1/2}(j+1/2)$ is the Dirac quantum number.
Then formula (\ref{wf_pnc1}) can be written as
    \begin{eqnarray}
 \label{wf_pnc1p}
|2^1S_0\ra \rightarrow |2^1S_0\ra +i\xi|2^3P_0\ra\,,
 \label{wf_pnc2p}
    \end{eqnarray}
where
\be \label{xi}
\xi = \frac{G_F}{2\sqrt{2}}\frac{Q_W}{E_{2^1S_0}-E_{2^3P_0}}
\int_{0}^{\infty}dr\,r^2
\rho_N(r)[g_{2p_{1/2}}f_{2s}-f_{2p_{1/2}}g_{2s}]\,.
\ee
The admixture of the $2^3P_0$  state enables the
$2^3S_1 - 2^3P_0$ transition, which is determined by the
electric-dipole amplitude. Since the electric-dipole transition
operator depends on the  gauge employed, the results
may differ in the different gauges, if the calculations are restricted
to a given approximation. The difference can be especially large for a
transition between the states having the same (or close)
 zeroth-order energies, as in the case under consideration.
In  Ref.~\cite{bra86} it was shown that using
the length gauge  in the calculation of the zeroth-order $2s - 2p_{1/2}$
transition amplitude in H-like ions is equivalent to accounting
for the one-loop QED corrections in the velocity-gauge
calculation, provided the transition energy in the length-gauge calculation
includes the corresponding corrections.
Let us show that accounting for the one-photon exchange and one-loop QED 
corrections to the  $2^3S_1 - 2^3P_0$ transition amplitude
in the velocity gauge can be performed equivalently within the zeroth-order
approximation in the length gauge by employing the transition energy
which includes the related corrections.

With this in mind, we consider first the evaluation of the
$2^3S_1 - 2^3P_0$  transition amplitude in the velocity gauge
to zeroth and first orders in $1/Z$.
The corresponding diagrams are presented in Figs. 1 and 2, respectively.
Formal expressions
for these diagrams can be derived using the two-time Green
function method \cite{sha02}. Such a derivation was considered in
detail in Ref.~\cite{ind04}. To simplify the analysis, we consider
the matrix elements of the electric-dipole transition operator
between the one-determinant wave functions,
\be \label{ua}
u_a(\bfx_1,\bfx_2)=\frac{1}{\sqrt{2}}\sum_{P}(-1)^P
\psi_{Pa_1}(\bfx_1)
\psi_{Pa_2}(\bfx_2)\,, \\
u_b(\bfx_1,\bfx_2)=\frac{1}{\sqrt{2}}\sum_{P}(-1)^P
\psi_{Pb_1}(\bfx_1)
\psi_{Pb_2}(\bfx_2)\,,
\label{ub}
\ee
where it is assumed that $a_1 = b_1 = 1s$,  $a_2 = 2s$,
$ b_2 = 2p_{1/2}$, $P$ is the permutation operator, and
$(-1)^P$ is the sign of the permutation.
Then, to zeroth order we obtain for the transition amplitude
\be
\tau^{(0)}&=&-\la b|[R_{\rm e}(1)+R_{\rm e}(2)]|a\ra =-\la b_1|R_{\rm e}(1)|a_1\ra \delta_{a_2\,b_2} -
\la b_2|R_{\rm e}(2)|a_2\ra \delta_{a_1\,b_1}\nonumber\\
&=& -\la 2p_{1/2}|R_{\rm e}|2s\ra\,.
\ee
Employing the identity
\be \label{id}
\balpha = i[H,{\bf r}]\,,
\ee
where $H$ is the one-electron
Dirac hamiltonian,
one obtains
\be \label{zeroth}
\tau^{(0)} =  i\frac{e}{\sqrt{2\omega (2\pi)^3}}
\la 2p_{1/2}|(\beps\cdot\br)|2s\ra
(\varepsilon_{2p_{1/2}}-\varepsilon_{2s})\,,
\ee
where  $\varepsilon_{2s}$ and $\varepsilon_{2p_{1/2}}$ are the
one-electron Dirac energies of the $2s$ and $2p_{1/2}$ states, respectively.
In particular, it follows that for the pure Coulomb field
($\varepsilon_{2s} = \varepsilon_{2p_{1/2}}$)
in the velocity gauge the zeroth-order $2^3S_1 - 2^3P_0$
transition amplitude is equal to zero.

The interelectronic-interaction
corrections, defined by the diagrams depicted in Fig. 2, consist of
irreducible and reducible contributions \cite{sha02,ind04}.
Since, to a good accuracy, these corrections can be treated with
the pure Coulomb field of the nucleus, in what follows, we restrict
our consideration to this approximation. Then, according to Ref. \cite{ind04}
we find that for the $2^3S_1 - 2^3P_0$
transition the reducible contribution vanishes.
As for the irreducible contribution,
it can be expressed as the sum \cite{ind04}
\be
\tau_{\rm irr} = \tau_{\rm irr}^{(a)} +\tau_{\rm irr}^{(b)},
\ee
where
\be \label{irred_a}
\tau_{\rm irr}^{(a)}&=&
  \frac{e}{\sqrt{2\omega (2\pi)^3}}
\sum_{P}(-1)^P
\Bigl\{
\sum_n^{\veps_{Pb_2}+\veps_{n}
\ne E_a^{(0)}}
\la Pb_1|(\beps\cdot\balpha)|n\ra \nonumber\\
&&\times
\frac{1}{E_a^{(0)}-\veps_{Pb_2}-\veps_n}
\la n Pb_2|I(\veps_{Pb_2}-\veps_{a_2})
|a_1 a_2\ra
\nonumber\\
&&+\sum_n^{\veps_{Pb_1}+\veps_{n}
\ne E_a^{(0)}}
\la Pb_2|(\beps\cdot\balpha)|n\ra
\frac{1}{E_a^{(0)}-\veps_{Pb_1}-\veps_n}
\la Pb_1 n|I(\veps_{Pb_1}-\veps_{a_1}
)|a_1 a_2\ra\Bigr\}
\,,
\ee
\be \label{irred_b}
\tau_{\rm irr}^{(b)}&=&
\frac{e}{\sqrt{2\omega (2\pi)^3}}
\sum_{P}(-1)^P
\Bigl\{
\sum_n^{\veps_{a_2}+\veps_{n}
\ne E_b^{(0)}}
\la Pb_1 Pb_2|I(\veps_{Pb_2}-\veps_{a_2})
|n a_2\ra \nonumber\\
&&\times
\frac{1}{E_b^{(0)}-\veps_{a_2}-\veps_n}
\la n|
(\beps\cdot\balpha)
|a_1\ra
\nonumber\\
&&
+\sum_n^{\veps_{a_1}+\veps_{n}
\ne E_b^{(0)}}
\la Pb_1 Pb_2|I(\veps_{Pb_1}-\veps_{a_1})
| a_1 n\ra
\frac{1}{E_b^{(0)}-\veps_{a_1}-\veps_n}
\la n|(\beps\cdot\balpha)
|a_2\ra\Bigr\}\,.
\ee
Here
$I(\omega)=e^2\alpha^{\mu}\alpha^{\nu}D_{\mu \nu}
(\omega)$,
     \begin{eqnarray}
            D_{\rho\sigma} (\omega,{\bf x}-{\bf y})=
              -g_{\rho\sigma}\int \;\frac{d{\bf k}}
                   {(2\pi)^{3}}\;
\frac{\exp{(i{\bf k}\cdot({\bf x}-{\bf y}))}}
                           {\omega^{2}-{\bf k}^{2}+i0}
         \label{e2e7}
     \end{eqnarray}
is the photon propagator in the Feynman gauge,
$\alpha^{\rho}\equiv \gamma^{0}\gamma^{\rho} =(1,\balpha)$,
$E_a^{(0)}=\veps_{a_1}+\veps_{a_2}$, and  $E_b^{(0)}=\veps_{b_1}+\veps_{b_2}$.
Taking into account that $E_a^{(0)}= E_b^{(0)}$
and using the identity (\ref{id}), we get
\be \label{irred_a1}
\tau_{\rm irr}^{(a)}&=&
  i\frac{e}{\sqrt{2\omega (2\pi)^3}}
\sum_{P}(-1)^P
\Bigl\{
\sum_n^{\veps_{n} \ne \veps_{Pb_1}}
\la Pb_1|(\beps\cdot\br)|n\ra
\la n Pb_2|I(\veps_{Pb_2}-\veps_{a_2})
|a_1 a_2\ra
\nonumber\\
&&+\sum_n^{\veps_{n} \ne \veps_{Pb_2}}
\la Pb_2|(\beps\cdot\br)|n\ra
\la Pb_1 n|I(\veps_{Pb_1}-\veps_{a_1}
)|a_1 a_2\ra\Bigr\}
\,,
\ee
\be \label{irred_b1}
\tau_{\rm irr}^{(b)}&=&
-i\frac{e}{\sqrt{2\omega (2\pi)^3}}
\sum_{P}(-1)^P
\Bigl\{
\sum_n^{\veps_{n} \ne \veps_{a_1}}
\la Pb_1 Pb_2|I(\veps_{Pb_2}-\veps_{a_2})
|n a_2\ra
\la n|
(\beps\cdot\br)
|a_1\ra
\nonumber\\
&&
+\sum_n^{\veps_{n} \ne \veps_{a_2}}
\la Pb_1 Pb_2|I(\veps_{Pb_1}-\veps_{a_1})
| a_1 n\ra
\la n|(\beps\cdot\br)
|a_2\ra\Bigr\}\,.
\ee
With the aid of the completeness condition
\be \label{comp}
\sum_n|n\ra \la n| = 1\,,
\ee
we find for the sum of the expressions (\ref{irred_a1}) and (\ref{irred_b1})
\be \label{first}
\tau_{\rm irr}=
  i\frac{e}{\sqrt{2\omega (2\pi)^3}}
\la 2p_{1/2}|(\beps\cdot\br)|2s\ra (\Delta E_b -\Delta E_a)\,,
\ee
where
\be
\Delta E_a = \sum_{P}(-1)^P\la Pa_1 Pa_2|I(\veps_{Pa_1}-\veps_{a_1})
| a_1 a_2\ra\,,
\ee
\be
\Delta E_b = \sum_{P}(-1)^P\la Pb_1 Pb_2|I(\veps_{Pb_1}-\veps_{b_1})
| b_1 b_2\ra\,
\ee
are the first-order interelectronic-interaction corrections to the initial
and final states, respectively. 
The same relation holds if one includes the one-loop QED corrections.
The corresponding proof, which
was given first in Ref. \cite{bra86}, is presented in the Appendix.
Summing up the zeroth- and first-order contributions yields
\be
\tau =  i\frac{e}{\sqrt{2\omega (2\pi)^3}}
\la 2p_{1/2}|(\beps\cdot\br)|2s\ra (E_b -E_a) = i\frac{\omega}{\sqrt{2\omega (2\pi)^3}}
\la 2p_{1/2}|(\beps\cdot\bd)|2s\ra \,,
\ee
where $\bd =e\br$ is the operator of electric-dipole moment, $E_a$  and $E_b$
are the total binding energies of the initial and final states, respectively. 
It is evident that similar equations can be derived involving two-electron
wave functions (\ref{wf_f}). Consequently, in what follows, we will use
the electric-dipole transition operator in the length gauge
\be
R^{(l)}_{\rm e} = - i\frac{\omega(\beps\cdot\bd) }{\sqrt{2\omega (2\pi)^3}}\,.
\ee
Substituting the two-electron wave function (\ref{wf_pnc1p}) into
Eq. (\ref{cross4}) and performing the calculation, we obtain for the PNC
contribution to the cross section
    \begin{eqnarray}
{\sigma_{\rm PNC}^{(2^3S_1\rightarrow 2^1S_0)}} =\frac{1}{9}\frac{\omega}{\Gamma_{2^3S_1} + \Gamma_{2^1S_0}}
2\lambda\xi \la 2p_{1/2}||\bd||2s\ra
\Bigl(\la 2s||\bmu||2s\ra - \la 1s||\bmu||1s\ra\Bigr)
\,,
 \label{cross41}
    \end{eqnarray}
where $ \la 2p_{1/2}||\bd||2s\ra $ is the reduced matrix element of the electric-dipole-moment
operator and $\lambda =\pm 1$ is the photon polarization.
Integrating over the angular variables in the reduced matrix elements yields
\be
\la ns||\bmu||ns\ra &=& -2e\sqrt{2/3}\int_0^{\infty}dr\,r^3g_{ns}(r)f_{ns}(r)\,,\\
 \la np_{1/2}||\bd||ns\ra &=& -e\sqrt{2/3}\int_0^{\infty}dr\,r^3(g_{np_{1/2}}(r)g_{ns}(r)
+ f_{np_{1/2}}(r)f_{ns}(r))\,.
 \ee
These integrals are easily evaluated employing the virial relations
for the Dirac equation (see, e.g., Refs. \cite{eps62,sha91,sha03}).
For the case of interest here, one derives
\be
2\int_0^{\infty}dr\,r^3(g_{2s}(r)f_{2s}(r) -g_{1s}(r)f_{1s}(r))&=& \gamma-\sqrt{(1+\gamma)/2}\,,\\
\int_0^{\infty}dr\,r^3(g_{2p_{1/2}}(r)g_{2s}(r)
+ f_{2p_{1/2}}(r)f_{2s}(r))&=&\frac{3(1+\gamma)\sqrt{1+2\gamma}}{2\alpha Z}\,,
\ee
where $\gamma = \sqrt{1-(\alpha Z)^2}$. Substituting these expressions into
Eqs. (\ref{cross40}), (\ref{cross41}) leads to
\be\label{sigma_0f}
\sigma_0^{(2^3S_1\rightarrow 2^1S_0)} = \frac{2}{27}\frac{\pi\alpha\omega}{\Gamma_{2^3S_1} + \Gamma_{2^1S_0}}
(\sqrt{2(1+\gamma)}-2\gamma)^2\,,
\ee
    \begin{eqnarray}
{\sigma^{(2^3S_1\rightarrow 2^1S_0)}}=
{\sigma_{0}^{(2^3S_1\rightarrow 2^1S_0)}}+
{\sigma_{\rm PNC}^{(2^3S_1\rightarrow 2^1S_0)}} =(1+\lambda\veps){\sigma_{0}^{(2^3S_1\rightarrow 2^1S_0)}}\,,
\label{cross42}
\ee
where
\be \label{eps_f}
\veps = 2\xi \frac{ \la 2p_{1/2}||\bd||2s\ra}
{\la 2s||\bmu||2s\ra - \la 1s||\bmu||1s\ra}= -2\xi\frac{3(1+\gamma)\sqrt{1+2\gamma}}
{\alpha Z(\sqrt{2(1+\gamma)} -2\gamma)}
    \end{eqnarray}
is a parameter which characterizes the relative value of the PNC effect.
The second term in the right-hand side of Eq. (\ref{cross42})
represents the PNC contribution, which changes the
sign under the replacement $\lambda \rightarrow -\lambda$.
The PNC parameter can also be represented as
\be
\veps=
-2\xi\sqrt{\frac{\Gamma_{2^3S_1} + \Gamma_{2^3P_0}}
{\Gamma_{2^3S_1} + \Gamma_{2^1S_0}}\;\frac{\sigma_0^{(2^3S_1\rightarrow 2^3P_0)}}{\sigma_0^{(2^3S_1\rightarrow 2^1S_0)}}}\,,
\ee
where $\sigma_0^{(2^3S_1\rightarrow 2^3P_0)}$ is the cross section of the resonant absorption
into the $2^3P_0$ state and $ \Gamma_{2^3P_0}$ is the total width of this state.

\section{Results and discussion}
\label{sec3}

The formulas (\ref{sigma_0f})-(\ref{eps_f}) allow one to evaluate
the cross section and the corresponding PNC effect. The most promising situation
for observing the PNC effect occurs in
cases where the levels $2^1S_0$  and   $2^3P_0$ are almost degenerate.
According to the most to-date elaborated calculations
 \cite{art05,kozh08b,yer06,kozh08a}
(see also the related table in Ref. \cite{mai09})
 such cases are gadolinium
($Z=64$) and thorium ($Z=90$), where the levels   $2^1S_0$  and   $2^3P_0$ are near to cross.
In case of Gd the energy interval amounts to -0.023(74) eV,
while in case of Th it is 0.44(40) eV \cite{kozh08a}. Since the
uncertainties are comparable to the energy differences,
to estimate the PNC effect we take the values 0.074 eV and 0.44 eV for the
$2^3P_0$ - $2^1S_0$ energy difference in Gd and Th, respectively.
The widths of the   $2^3S_1$ and $2^1S_0$ states, which enter
formula (\ref{sigma_0f}),
are mainly defined by
the one-photon M1 and two-photon E1E1 transitions, respectively.
We evaluate the decay rate of the M1-transition $2^3S_1\rightarrow 1^1S_0$ employing
the transition energy taken from Ref.~\cite{art05}. The interelectronic-interaction
corrections to the transition amplitude are calculated to first order in $1/Z$
within a systematic QED approach (see for details Ref.~\cite{ind04}).
As the result, we obtain decay rates
$w_{\rm M1}^{(2^3S_1\rightarrow 1^1S_0)} = 2.301\times 10^{12}$ s$^{-1}$ for Gd and
$w_{\rm M1}^{(2^3S_1\rightarrow 1^1S_0)} = 9.470\times 10^{13}$ s$^{-1}$ for Th.
These values are in a fair agreement with those from Refs.~\cite{ind04,joh95,and09}.
The two-photon decays $2^3S_1\rightarrow 1^1S_0$ and $2^1S_0\rightarrow 1^1S_0$
are calculated in the length gauge with the transition energies taken from Ref.~\cite{art05}.
The interelectronic-interaction effects are approximately accounted for
by means of a Kohn-Sham potential. The calculated transition rates are
$w_{2\gamma}^{(2^3S_1\rightarrow 1^1S_0)} = 8.74\times 10^{ 8}$ s$^{-1}$,
$w_{2\gamma}^{(2^1S_0\rightarrow 1^1S_0)} = 9.04\times 10^{11}$ s$^{-1}$  in case of Gd and
$w_{2\gamma}^{(2^3S_1\rightarrow 1^1S_0)} = 2.07\times 10^{10}$ s$^{-1}$,
$w_{2\gamma}^{(2^1S_0\rightarrow 1^1S_0)} = 6.25\times 10^{12}$ s$^{-1}$  in case of Th.
These values together with the E1E1 channel include also higher multipole contributions,
such as M1M1 etc. In case of Th the dominant E1E1 decay channel yields
$w_{\rm E1E1}^{(2^3S_1\rightarrow 1^1S_0)} = 1.62\times 10^{10}$ s$^{-1}$ and
$w_{\rm E1E1}^{(2^1S_0\rightarrow 1^1S_0)} = 6.25\times 10^{12}$ s$^{-1}$.
It is worth noticing that for the $2^3S_1$ state the higher multipoles contribute
up to 20\% to the total two-photon decay rate.
Comparing the E1E1 decay rates with the results of Ref.~\cite{der97}, we find
an excellent agreement for the $2^3S_1$ state and a slight deviation for the $2^1S_0$ state,
which is mainly due to employing the more accurate transition energies
in our calculations.
Finally, the total widths are
$\Gamma_{2^3S_1} = 1.515$ meV, $\Gamma_{2^1S_0} = 0.595$ meV in case of Gd, and
$\Gamma_{2^3S_1} = 62.35$ meV, $\Gamma_{2^1S_0} = 4.11$ meV in case of Th.
The results of the calculations of the PNC effect by formulas  (\ref{sigma_0f})-(\ref{eps_f})
for Gd and Th are presented in Table I.

As one can see from the table, in both Gd and Th cases the PNC effect amounts
to about 0.05\%, which is  a rather large value for parity-violation experiments.
What is more, one may expect that the PNC effect can be further increased,
at least, by an order of magnitude by choosing proper isotopes,
provided the $2^1S_0 - 2^3P_0$ energy difference
is known to a higher accuracy. With the current experimental
techniques \cite{bra08}, accurate measurements of the
difference considered seem feasible.

Because of a large transition energy ($>$ 100 eV),
until recently  the experimental scenarios with the laser-induced
 $2^3S_1 - 2^1S_0$ transition in heavy He-like ions were  far
from being possible. However, the situation has changed in view of the
very significant progress in X-ray laser development.
Such lasers will be available in the near future with a high repetition rate
\cite{hei06}. 
Already now, there is a first X-ray laser available at 
the heavy ion facility GSI (PHELIX facility)  
where photon energies of up to 200 eV have been reached \cite{zim08,Bag09}.
As an alternative scenario, the excitation energy
can be obtained by counter-propagating the
ultraviolet laser beam with the photon energy in the range from 4 to 10 eV
and the He-like ion beam with the energy  up to 10.7 GeV/u,
which will be available  at the FAIR  facility in Darmstadt  \cite{Bac00,Sto06}.
The population of the  $2^1S_0$ level
can be measured by observing the 2E1 decay to the ground state.
In the second scenario, due to the strong Lorentz  boosting,
the decay photons are emitted at the forward direction,
that considerably simplifies their detection.

The next problem to be addressed is the preparation of
ions in the $2^3S_1$ state
that is required in both scenarios considered. As follows
from the study presented in Ref. \cite{sto98,rza06}, in collisions with gas atoms
one can produce selectively both
the $2^1S_0$ state  and  the $ 2^3S_1$ one \cite{Trotsenko}.
However, it would be of great importance to populate exclusively only the $2^3S_1$ state.
The only way to accomplish this 
is to form first the doubly excited $(2s2p_{1/2})_0$  state
via dielectronic recombination of an electron with a H-like ion.
Since the main decay channel of the $(2s2p_{1/2})_0$ state is
the transition into the $2^3S_1$ state, this enables
selective production of ions in the $2^3S_1$ state.

The PNC effect is to be measured by counting the intensity
difference in the 2E1 decay of the  $2^1S_0$ state for polarizations
$\lambda=\pm 1$.
The background emission can be separated by switching off the laser
light. Changing the photon energy allows one to eliminate the interference
with a non-resonant transition via the  $2^3P_0$ state, which could
also be evaluated to a good accuracy if necessary.
Moreover, since the 2E1
emission can be measured relative to the intensity of the M1 X-ray line
(decay of the $2^3S_1$ state), such an
experimental scenario appears to be quite realistic.

\section{Conclusion}
In this paper we have studied the PNC effect
with  laser-induced
$2^3S_1$- $2^1S_0$ transition in heavy heliumlike ions.
A simple analytical formula for  the
photon-absorption cross section derived enables easy
evaluation of the PNC effect for ions nearby $Z=64$ and $Z=90$, where
the effect is strongly enhanced due to near-crossing of
the opposite-parity  $2^1S_0$ and  $2^3P_0$ levels.
The calculations performed showed that the effect can amount
to about 0.05\% and even bigger for the ions of interest.
Prospects for the corresponding PNC experiments have been discussed.
It is found that the desired photon energy can be achieved either
by X-ray lasers that are presently
getting developed at GSI (PHELIX project) as well as 
at the Helmholtz-Institute in Jena \cite{46}
or by counter-propagating the
ultraviolet laser beam
and the He-like ion beam at the FAIR  facility in Darmstadt .

\section*{Acknowledgments}
Valuable discussions with Andrey Surzhykov are gratefully acknowledged.
This work was supported by DFG (Grants No. 436RUS113/950/0-1 and VO 1707/1-1),
by GSI, by RFBR (Grant No. 10-02-00450), by the Ministry of
Education and Science of Russian
Federation (Program for Development of Scientific Potential of
High School, Grant No. 2.1.1/1136; Program ``Scientific and
pedagogical specialists for innovative Russia'', Grant No. P1334), and by
the ExtreMe Matter Institute EMMI in the framework of the Helmholtz Alliance HA216/EMMI.
V.M.S. acknowledges financial support by the Alexander von Humboldt Foundation.

\newpage

\appendix*
\section{QED corrections to the transition amplitude}
The one-electron QED corrections to the   $2^3S_1 - 2^3P_0$
transition amplitude
are determined by the corresponding contributions to
the $2s-2p_{1/2}$ amplitude in one-electron ions
as defined by the diagrams shown in Fig. 3.
Formal expressions for these corrections are almost the same as for the
corresponding corrections to the emission amplitude \cite{sha02}.
Let us consider the one-loop self-energy correction.
According to formulas provided in Ref. \cite{sha02}, it is given by the
sum of the irreducible, reducible, and vertex contributions.
For the electron interacting with pure Coulomb field together
with the dipole approximation $\exp{(i\bfk\cdot \bfx)} \rightarrow 1$,
the reducible contribution vanishes. The irreducible
contribution is given by
\be \label{se_irr}
\tau^{\rm (SE)}_{\rm irr} &=&  -\la 2p_{1/2}|R|\xi_{2s}\ra  -\la \xi_{2p_{1/2}}|R|2s\ra\nonumber\\
&=& \frac{e}{\sqrt{2\omega (2\pi)^3}}[ \la 2p_{1/2}|(\beps\cdot\balpha)|\xi_{2s}\ra
+\la \xi_{2p_{1/2}}|(\beps\cdot\balpha)|2s\ra]\,,
\ee
where
\be
|\xi_a\ra = \sum_{n}^{n\ne a}\frac{|n\ra \la  n|\Sigma(\veps_a)|a\ra}{\veps_a-\veps_n}\,,
\ee
\be
\la \xi_b| = \sum_{n}^{n\ne b}\frac{ \la  b|\Sigma(\veps_b)|n\ra\la n|}{\veps_b-\veps_n}\,,
\ee
and
\be
\la a|\Sigma(\veps)|b\ra=\frac{i}{2\pi} \int_{-\infty}^{\infty}
d\omega \;
\sum_{n}\frac{\la an|e^2\alpha^{\rho}\alpha^{\sigma}
D_{\rho \sigma}(\omega)|n b\ra}{\veps-\omega-\veps_n(1-i0)} \,.
\ee
By means of the identity (\ref{id}) and the completeness relation
(\ref{comp}), we obtain
\be
\tau^{\rm (SE)}_{\rm irr} &=& i \frac{e}{\sqrt{2\omega (2\pi)^3}}
\Bigl\{\la  2p_{1/2}|(\beps\cdot\br)|2s\ra
(\la  2p_{1/2}|\Sigma(\veps_{2p_{1/2}})| 2p_{1/2}\ra -
\la  2s|\Sigma(\veps_{2s})| 2s\ra)\nonumber\\
&&+\la  2p_{1/2}|[(\beps\cdot\br),\Sigma(\veps_{2s})]|2s\ra \Bigr\} \,.
\ee
For the vertex contribution one derives
\be
\tau^{\rm (SE)}_{\rm ver} &=& -e^2\frac{i}{2\pi}\int_{-\infty}^{\infty}\, d\omega\;
\int\,\frac{d\bfk}{(2\pi)^3}\;\frac{1}{ \omega^{2}-{\bf k}^{2}+i0}
\sum_{n_1,n_2}\la 2p_{1/2}|\alpha^{\rho}\exp{(i\bfk\cdot \bfy)}|n_1\ra\nonumber\\
&&\times \frac{1}{\veps_{2p_{1/2}}-\omega-\veps_{n_1}(1-i0)}
\la n_1|
 \frac{e(\beps\cdot\balpha)}{\sqrt{2\omega (2\pi)^3}}
|n_2\ra \frac{1}{\veps_{2s}-\omega-\veps_{n_2}(1-i0)}\nonumber\\
&&\times
\la n_2|\alpha_{\rho}\exp{(-i\bfk\cdot \bfx)}|2s\ra\,.
\ee
Transforming
\be
\frac{1}{\veps_{2p_{1/2}}-\omega-\veps_{n_1}(1-i0)} \frac{1}{\veps_{2s}-\omega-\veps_{n_2}(1-i0)}
\hspace{5cm}\nonumber\\
=\frac{1}{\veps_{n_1}-\veps_{n_2}}\Bigl(\frac{1}{\veps_{2p_{1/2}}-\omega-\veps_{n_1}(1-i0)} -
\frac{1}{\veps_{2s}-\omega-\veps_{n_2}(1-i0)}\Bigr)\,,
\ee
\be
\la n_1|(\beps\cdot\balpha)|n_2\ra =
i\la n_1|[H,(\beps\cdot{\bf r})]|n_2\ra =i(\veps_{n_1}-\veps_{n_2})\la n_1|(\beps\cdot{\bf r})|n_2\ra\,,
\ee
we get
\be
\tau^{\rm (SE)}_{\rm ver} &=& -\frac{e}{\sqrt{2\omega (2\pi)^3}}
 e^2\frac{i}{2\pi}\int_{-\infty}^{\infty}\, d\omega\;
\int\,\frac{d\bfk}{(2\pi)^3}\;\frac{1}{ \omega^{2}-{\bf k}^{2}+i0}
\sum_{n_1,n_2}^{\veps_{n_1} \ne \veps_{n_2}}\la 2p_{1/2}|\alpha^{\rho}\exp{(i\bfk\cdot \bfy)}|n_1\ra\nonumber\\
&&\times
i\Bigl(\frac{1}{\veps_{2p_{1/2}}-\omega-\veps_{n_1}(1-i0)}-
\frac{1}{\veps_{2s}-\omega-\veps_{n_2}(1-i0)}\Bigr)
\la n_1|(\beps\cdot\br)|n_2\ra \nonumber\\
&&\times
\la n_2|\alpha_{\rho}\exp{(-i\bfk\cdot \bfx)}|2s\ra\nonumber\\
 &=& -i\frac{e}{\sqrt{2\omega (2\pi)^3}}
 e^2\frac{i}{2\pi}\int_{-\infty}^{\infty}\, d\omega\;
\int\,\frac{d\bfk}{(2\pi)^3}\;\frac{1}{ \omega^{2}-{\bf k}^{2}+i0}
\Bigl\{
\sum_{n_1}\la 2p_{1/2}|\alpha^{\rho}\exp{(i\bfk\cdot \bfy)}|n_1\ra
\nonumber\\
&&\times
\frac{1}{\veps_{2p_{1/2}}-\omega-\veps_{n_1}(1-i0)}\Bigl[
\la n_1|(\beps\cdot\bfx)\alpha_{\rho}\exp{(-i\bfk\cdot \bfx)}|2s\ra\nonumber\\
&&-\sum_{n_2}^{\veps_{n_1} = \veps_{n_2}}\la n_1|(\beps\cdot\bfr)|n_2\ra\la n_2|
\alpha_{\rho}\exp{(-i\bfk\cdot \bfx)}|2s\ra\Bigr]\nonumber\\
&&-\sum_{n_2}\Bigl[\la 2p_{1/2}|(\beps\cdot\bfy)\alpha^{\rho}\exp{(i\bfk\cdot \bfy)}|n_2\ra
\nonumber\\
&&-\sum_{n_1}^{\veps_{n_1} = \veps_{n_2}}\la 2p_{1/2}|\alpha^{\rho}\exp{(i\bfk\cdot \bfy)}|n_1\ra
\la n_1|(\beps\cdot\bfr)|n_2\ra\Bigr]\nonumber\\
&&\times
\frac{1}{\veps_{2s}-\omega-\veps_{n_2}(1-i0)}\la n_2|\alpha_{\rho}\exp{(-i\bfk\cdot \bfx)}|2s\ra\Bigr\}\nonumber\\
&=& -i \frac{e}{\sqrt{2\omega (2\pi)^3}}
\la  2p_{1/2}|[(\beps\cdot\br),\Sigma(\veps_{2s})]|2s\ra \,.
\ee
The sum of both irreducible and vertex contributions yields \cite{bra86}
\be
\tau^{\rm (SE)}_{\rm tot} &=& i \frac{e}{\sqrt{2\omega (2\pi)^3}}
\la  2p_{1/2}|(\beps\cdot\br)|2s\ra
(\la  2p_{1/2}|\Sigma(\veps_{2p_{1/2}})| 2p_{1/2}\ra -
\la  2s|\Sigma(\veps_{2s})| 2s\ra) \,.
\ee
A similar equation can be derived for the vacuum-polarization contribution.

\begin{table}
\caption{The zeroth-order cross section $\sigma_{0}^{(2^3S_1\rightarrow 2^1S_0)}$,
the PNC correction $\sigma_{\rm PNC}^{(2^3S_1\rightarrow 2^1S_0)}$, and the parameter
$\veps$, defined by Eq. (\ref{cross42}),
for the laser-induced $2^3S_1\rightarrow 2^1S_0$ transition in He-like Gd and Th.
The  $2^3S_1$ - $2^1S_0$ transition energies  are
taken from Ref. \cite{art05}, while the
 $2^3P_0$ - $2^1S_0$ energy difference is chosen as discussed in the text.}
\vspace{0.2cm}
\label{tab0}
\begin{ruledtabular}
\begin{tabular} {c|c|c|c|c|c}
\hline
Ion & $(E_{2^3P_0} - E_{2^1S_0})$ [eV]   & $(E_{2^1S_0} - E_{2^3S_1})$ [eV] &
$\sigma_{0}^{(2^3S_1\rightarrow 2^1S_0)}$ [barn] &   $\sigma_{\rm PNC}^{(2^3S_1\rightarrow 2^1S_0)}$ [barn] &
$\veps$ \hspace*{0.115cm} \\
\hline
$^{158}$Gd$^{62+}$ & 0.074(74) & 114.0 & 4084.1 & $\pm$ 2.1 &  -0.00051 \hspace*{0.115cm} \\
\hline
$^{232}$Th$^{88+}$ & 0.44(40) & 240.1 & 1217.6 & $\pm$ 0.6 & -0.00053 \hspace*{0.115cm} \\
\hline
\end{tabular}
\end{ruledtabular}
\end{table}

\begin{figure}
\includegraphics{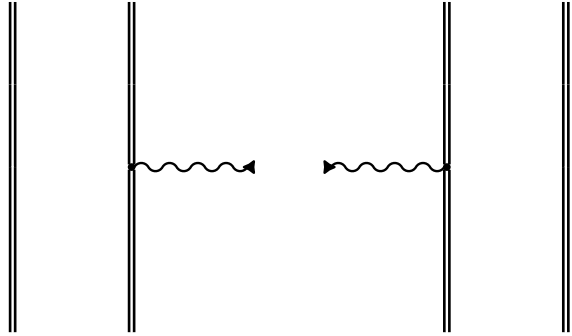}
\caption{The absorption of a photon by a
heliumlike ion to the zeroth-order approximation (noninteracting electrons).}
\end{figure}

\begin{figure}
\includegraphics{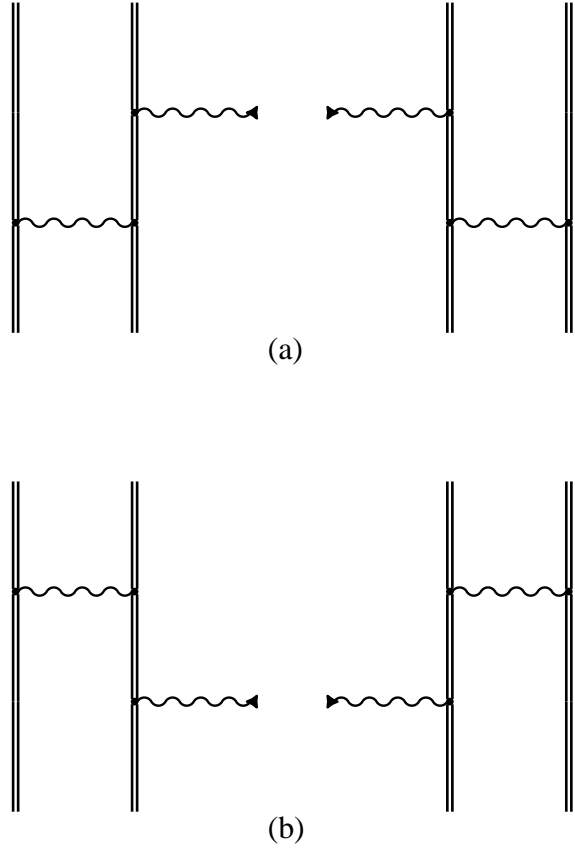}
\caption{One-photon exchange corrections
to the absorption of a photon by a heliumlike ion.}
\end{figure}

\begin{figure}
\includegraphics{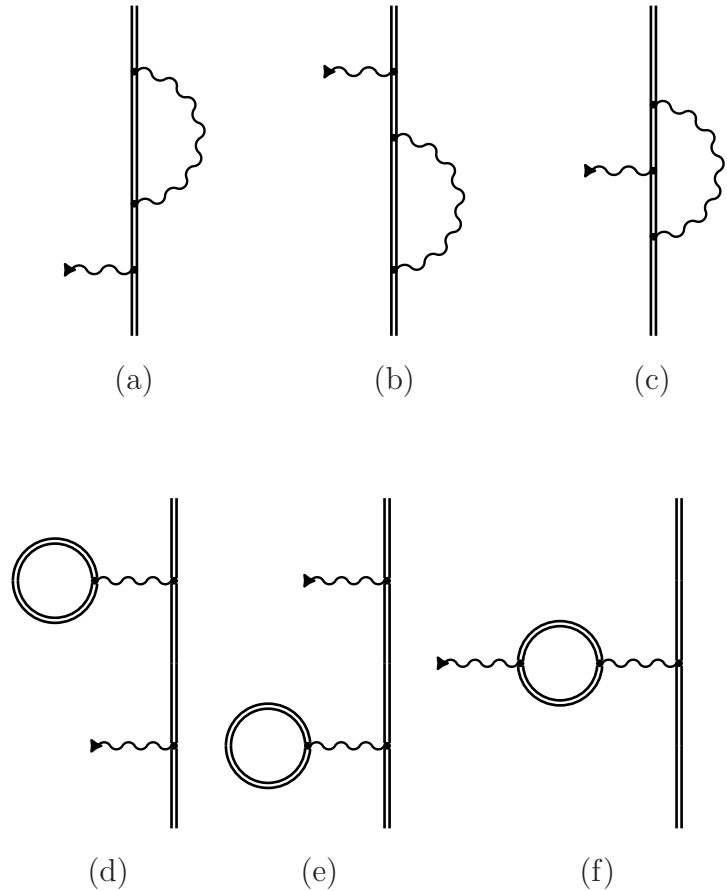}
\caption{One-loop QED corrections to the photon absorption.}
\end{figure}

\end{document}